# Possible temperature control DC switch effect between two superconductors

Tian De Cao

*Department of physics, Nanjing University of Information Science & Technology, Nanjing 210044, China*

**Abstract**

The lifetime of an electron pair could not be unlimited long, on the basis of this, we suggest a model. The model means that the movements of charge carriers in a superconductor should have three forms: the single-electron movement, the single-pair movement, and the revolving around the mass center of two electrons in a pair. Thus the current in a superconductor has three possible parts. Similarly, there should be three possible effects in a SIS junction: the tunneling of single electron, the tunneling of single pair, and the pair-forming following the pair-breaking. This paper will discuss these problems and present a possible temperature control DC switch effect between two superconductors.

**1. Introduction**

To explain the superconductivity, Cooper presented the electron pair which has zero momentum and zero spin [1]. The Cooper pairs should be most stable and their boson feature can explain this. However, there should be non-zero momentum pairs, in spite of whether there was supercurrent or not in a superconductor. Moreover, the revolving around their mass center by the two electrons in a pair may contribute to the current. That is to say, the current in a superconductor should have three origins, due to the movements of pairs, due to the movements of electrons, and due to the revolving. We will find that each of these three kinds of currents has its particular feature, especially in the tunneling process. Correspondingly, there should be three possible effects in a SIS junction, two of them have been reported and observed, the tunneling of the single electron and the Josephson effect [2]. However, the Josephson effect has been explained as the tunneling of the electron pairs, this may be wrong. We find that the Josephson effect is due to the pair-forming following pair-breaking. Thus the tunneling of the electron pairs has not been reported in fact. In this paper, three kinds of current in a superconductor and three kinds of tunneling in a SIS junction are discussed.

**2. Model**

We will discuss the BCS superconductors, for simplicity, while this discussion can be extended to other superconductors. The observable particles in a superconductor should be around the Fermi surface, thus the charge carriers are electrons (or holes) and superconducting pairs. Thus the Hamiltonian of electron systems should be written in three parts, $H = H_e^{eff} + H_{pair} + H_{e-pair}$. Because the lifetime of a pair is not infinite, $H_{e-pair}$ must include two terms similar to $c_{k\sigma}^+ c_{-k+q\bar{\sigma}}^+ a_q$ and $c_{-k+q\bar{\sigma}} c_{k\sigma} a_q^+$, the former expresses the pair-breaking, the latter expresses the pair-forming, thus we consider this Hamiltonian [3]

$$H = H_e^{eff} + H_{pair} + H_{e-pair} \qquad (1)$$

$$H_e^{eff} = \sum_{k,\sigma} \xi_k c_{k\sigma}^+ c_{k\sigma} - \frac{1}{2} V \sum_{\substack{k,k',q \\ \sigma,\sigma'}} c_{k+q\sigma}^+ c_{k\sigma} c_{k'-q\sigma'}^+ c_{k'\sigma'}$$

$$H_{pair} = \sum_q \omega_q a_q^+ a_q$$

$$H_{e-pair} = \sum_{k,q,\sigma} v_{k,q} c_{k\sigma}^+ c_{\bar{k}+q\bar{\sigma}}^+ a_q + \sum_{k,q,\sigma} v_{k,q}^* c_{\bar{k}+q\bar{\sigma}} c_{k\sigma} a_q^+$$

for the charge carriers around the Fermi level of a spin-singlet superconductor. The electron operator $c_{k\sigma}$ and the boson operator $a_q$ describe the electrons and superconducting pairs, respectively. We have denoted wave vector $\vec{k}$ as $k$, $k \equiv \vec{k}$. The electron Hamiltonian $H_e^{eff}$ describes the unpaired electrons (but they include the ones in the pair-forming and the pair-breaking process) and it can reproduce the Green's functions and the unusual functions of BCS theory, while $H_{pair}$ describes the paired electrons. Someone may believe $H = H_e^{eff}$ for the BCS superconductors, yes or no? In the past, what many works handle is just the so-called "pair-forming process" and "pair-breaking process" (not the "paired state") in the aspect of superconductivity, in all these works, we can take $H = H_e^{eff}$. In describing superconductivity, we have used the unusual functions [4] which describe the zero momentum motion of each pair, and they belong to the "unusual propagating functions" which only describe "pair-forming process" (two electrons in a pair are produced at $\tau$ time and at $\tau'$ time respectively) and "pair-breaking process". However, if the movement of each pair could not be neglected, the Hamiltonian should include $H_{pair}$ and $H_{e-pair}$ as shown in Eq.(1).

In concerned problems, we may encounter with the Green's function $G(k\sigma, \tau - \tau') = - <T_\tau c_{k\sigma}(\tau) c_{k\sigma}^+(\tau')>$ and the unusual functions $F(k\sigma, \tau - \tau')$ $=<T_\tau c_{\bar{k}\bar{\sigma}}(\tau)c_{k\sigma}(\tau')>$ and $F^+(k\sigma,\tau-\tau')\ =<T_\tau c^+_{k\sigma}(\tau)c^+_{\bar{k}\bar{\sigma}}(\tau')>$. These functions are determined by $H^{eff}$ but affected by $H_{e-pair}$, this is because the electron states should be affected by the pair-forming and the pair-breaking. Here we have noted what a function describes and what a Hamiltonian includes. When there is current in a superconductor, both $\xi_k$ and $\omega_q$ are correlated with the magnetic vector potential, what are the features of the current? Generally, the current (density) could be written in $\vec{j} = \vec{j}^{\,pair} + \vec{j}^{\,e} + \vec{j}\,'$, $\vec{j}^{\,pair}$ is due to the movements of pairs, $\vec{j}^{\,e}$ is due to the movements of electrons, and $\vec{j}\,'$ is due to that the two electrons in a pair revolve around their mass center. Because $\vec{j}^{\,pair}$ is correlated to the operators $c_{\bar{k}\bar{\sigma}}c_{k+q\sigma}$ and $c^+_{k+q\sigma}c^+_{\bar{k}\bar{\sigma}}$ [5], thus $\vec{j}^{\,pair}$ is due to the movement of the mass center of a pair. If $\vec{j}^{\,e}$=0, the resistance of the superconductor is zero. It is well known that London presented a relation between suppercurrent density and magnetic vector potential [6], Pippard suggested a non-local relation [7], and the similar relation seems having been derived with the BCS theory (and with some strongly correlated models). However, the relation between the current and the superconducting phase could not be derived from the BCS theory (and other theories), why? We suggest that the current from the BCS theory is just $\vec{j}\,'$ which is because the two electrons in a pair revolve around their mass center, and this could be found in its derivation because $\vec{j}\,'$ is correlated to the pair operators $c_{\bar{k}\bar{\sigma}}c_{k\sigma}$ and $c^+_{k\sigma}c^+_{\bar{k}\bar{\sigma}}$ (they correspond to zero momentum of a pair). We have established the equation of $\vec{j}^{\,pair}$ associated with both superconducting phase and magnetic vector potential. It is shown that both $\vec{j}^{\,pair}$ and $\vec{j}\,'$ have the zero-resistance effect and the Meissner effect. Which is important for $\vec{j}^{\,pair}$ and $\vec{j}\,'$? This has to be investigated.

## 3. Three kinds of tunneling currents

To discuss the tunneling problem, we define the electron number and the boson number in $N^e = \sum_{k,\sigma} c^+_{k\sigma} c_{k\sigma}$ and $N^{pair} = \sum_q a^+_q a_q$ respectively, it is easy to find $[H, N^e] \neq 0$, $[H, N^{pair}] \neq 0$, but $[H, N^e + 2N^{pair}] = 0$. This is consistent with the total electron number conservation. This also means that if we will discuss their statistics, the chemical potential of electrons (which are not in the superconducting state) is $\mu$, and the chemical potential of bosons is $2\mu$.

To find the main features of the tunneling problem, the SIS tunneling Hamiltonian is taken in the form

$$H_T = \sum_{k,p,\sigma}(T_{kp} c^+_{k\sigma} d_{p\sigma} + T^*_{kp} d^+_{p\sigma} c_{k\sigma}) + \sum_{k,p}(\tau_{kp} a^+_k b_p + \tau^*_{kp} b^+_p a_k) \tag{2}$$

The electron operator on the left is expressed in terms of one set of operators $c_{k\sigma}$ and those on the right by another set $d_{p\sigma}$, and the boson operator on the left is expressed in terms of one set of operators $a_k$ and those on the right by another set $b_p$. Following the derivation in some books, we take $H_L = H_L^{eff} + \sum_q \omega_{Lq} a^+_q a_q + \sum_{k,q,\sigma} v_{Lk,q} c^+_{k\sigma} c^+_{\bar{k}+q\bar{\sigma}} a_q + \sum_{k,q,\sigma} v^*_{Lk,q} c_{\bar{k}+q\bar{\sigma}} c_{k\sigma} a^+_q$ and $H_R = H_R^{eff} + \sum_q \omega_{Rq} b^+_q b_q + \sum_{k,q,\sigma} v_{Rk,q} d^+_{k\sigma} d^+_{\bar{k}+q\bar{\sigma}} b_q + \sum_{k,q,\sigma} v^*_{Rk,q} d_{\bar{k}+q\bar{\sigma}} d_{k\sigma} b^+_q$, and we find $[H_L, N^e_L + 2N^{pair}_L] = 0$ and $[H_R, N^e_R + 2N^{pair}_R] = 0$. In the calculation below, we define $K_L = H_L - \mu_L N^{total}_L$ and $K_R = H_R - \mu_R N^{total}_R$. The current is calculated by $I = -e < \dot{N}^e_L + 2\dot{N}^{pair}_L >$ under the applied voltage $V$. Because $\dot{N}^e_L + 2\dot{N}^{pair}_L = i\sum_{k,p,\sigma}(-T_{kp} c^+_{k\sigma} d_{p\sigma} + T^*_{kp} d^+_{p\sigma} c_{k\sigma}) + i2\sum_{k,p}(-\tau_{kp} a^+_k b_p + \tau^*_{kp} b^+_p a_k)$ and the time development of $d_{k\sigma}$ operator is governed by $d_{k\sigma}(t) = e^{iK_R t} d_{k\sigma} e^{-iK_R t}$ and so on, as the usual case in linear approximation, we arrive at $I = I_s + I_J = I^e + I^{pair} + I_J$ and

$$I^e = e \int_{-\infty}^{t} dt' \{ e^{ieV(t-t')} <[A^+(t), A(t')]> - e^{-ieV(t-t')} <[A(t), A^+(t')]> \} \quad (3)$$

$$I^{pair} = e \int_{-\infty}^{t} dt' \{ e^{i2eV(t-t')} <[B^+(t), B(t')]> - e^{-i2eV(t-t')} <[B(t), B^+(t')]> \} \quad (4)$$

$$I_J = e \int_{-\infty}^{t} dt' \{ e^{ieV(t+t')} <[A^+(t), A^+(t')]> - e^{-ieV(t+t')} <[A(t), A(t')]> \} \quad (5)$$

where $A(t) = \sum_{k,p,\sigma} T^*_{kp} d^+_{p\sigma}(t) c_{k\sigma}(t)$ and $B(t) = \sum_{k,p} 2\tau^*_{kp} b^+_p(t) a_k(t)$. We have noted $<B^+(t), B^+(t')]> = 0$ and $<[B(t), B(t')]> = 0$. Eq.(3) describes the tunneling of single electron, it is the well-known one, and one of its features is $I^e \propto \theta(eV - 2\Delta)$ for BCS superconductors. Eq.(5) describes the Josephson effect, its main feature is $I_J \sim I_J(eV)\sin(\omega t + \phi)$. Eq.(5) includes the functions $<T_\tau c_{\bar{k}\bar{\sigma}}(\tau) c_{k\sigma}(\tau')>$ and $<T_\tau d^+_{p\sigma}(\tau) d^+_{\bar{p}\bar{\sigma}}(\tau')>$ which describe the pair-breaking process and the pair-forming process, these processes coupled by $T_{kp}$, and these processes express the pair-forming following pair-breaking. In this paper, we discuss the "new term" $I^{pair}$, and we arrive at

$$I^{pair} = 2e \sum_{k,p} |\tau_{kp}|^2 \int_{-\infty}^{+\infty} \frac{d\varepsilon}{2\pi} A^{pair}_R(p,\varepsilon) A^{pair}_L(k,\varepsilon + 2eV)[n_B(\varepsilon) - n_B(\varepsilon + 2eV)] \quad (6)$$

where we introduced the boson Green's function $D_L(q, \tau - \tau') = -<T_\tau a_q(\tau) a^+_q(\tau')>$, the boson spectral function $A^{pair}_L(q,\Omega) = -2\text{Im} D^{ret}_L(q,\Omega)$, and so on.

**4. Possible DC switch effect**

Have one already observed $I^{pair}$ in experiments? There are two possibilities. One, $I^{pair} \neq 0$ but experimenters regard it as "noise current". Two, experimenters indeed observed $I^{pair}=0$ in some experiments. In the second case, it is because $\tau_{kp}=0$. $\tau_{kp} \neq 0$ should require that the junction is very thin. Moreover, the calculation of Eq. (6) is complex than Eq.(3), this is because $\tau_{kp}$ should obviously depend on the wave vectors $k$ and $p$, and this dependence has to be determined. One of our conjectures is that $\tau_{kp}$ decreases with increasing wave vectors due to the pair-breaking effect. In another aspect, Eq.(6) shows that $I^{pair}$ depends on the Boson-Einstein distribution, thus $I^{pair}$ increases with decreasing temperature. Moreover, $\tau_{kp}$ may be larger in some SIS junction when the insulator consists of a particular material.

If $I^{pair} \neq 0$, it is not difficult to find

$$I^{pair} = \frac{2eV}{k_B T} f(2eV, T) \qquad (7)$$

for $eV < k_B T$. This is different from the single electron tunneling current, because of $I^e \propto \theta(eV - 2\Delta)$ for the BCS superconductors. If $I^{pair}$ will be observed, what are the possible significances? One, some technology will be improved. Because $T \ll T_c$ favors $I^{pair} \neq 0$, the superconductors in a SIS junction should be high temperature superconductors, but the high temperature superconductivity is usually d-wave symmetry (not the BCS superconductors), thus the forming of an effective junction is not easy. Two, we will take some progress in understanding superconductivity. For example, a whole microscopic equation linked the supercurrent to both phase and magnetic vector potential is first found in our theory, the microscopic explanation of superconducting phase is first done, and so on [5]. If Eq.(7) is confirmed with experiments, the theory is also examined.

Three, because $I^e = 0$ for $eV < 2\Delta$, the DC current is cut off. However, when the temperature is so low that $I^{pair} \neq 0$, the DC current is connected. This is the so-called temperature control switch, and this may be applied in some technologies.

**5. Summary and discussion**

Since there are the processes of the pair-forming and the pair-breaking in a superconductor, their effects could not be neglected in tunneling problems, thus we present the Hamiltonian (1) of a BCS superconductor, and it can be extended to other superconductors if $H_e^{eff}$ will be improved. This necessary gives three kinds of tunneling currents as shown in Eqs (3)-(5). The Josephson current in Eq.(5) is correlated with the functions $<T_\tau c_{\bar{k}\bar{\sigma}}(\tau)c_{k\sigma}(\tau')>$ and $<T_\tau d^+_{p\sigma}(\tau)d^+_{\bar{p}\bar{\sigma}}(\tau')>$ which describe the pair-breaking process and the pair-forming process, and it is related to the zero-momentum pair. The tunneling of pairs is described by Eq.(4), and it gives possible temperature control DC switch effect which could occur in some SIS junctions.

**References**

[1] Cooper L N, 1956 Phys. Rev. **104** 1189.

[2] Josephson B D, 1962 Phys. Lett. **1**, 251.

[3] Cao T D, arXiv:1010.3896.

[4] Mahan G D, Many-particle physics, Chap. 9, p.778 (Plenum Press, New York, 1990).

[5] Cao T D, arXiv:1107.3282.

[6] London E, Superfluids, Vol.1(Wiley, New York,1954).

[7] Pippard A B, 1953 Proc.Roy.Soc., **A216** 547.